\documentclass[aps,prl,twocolumn,superscriptaddress,showpacs,preprintnumbers,amsmath,amssymb]{revtex4}
\usepackage{graphicx} 
\usepackage{dcolumn}  

\graphicspath{{ps}}

\def\be{\begin{equation}}
\def\ee{\end{equation}}

\newcommand{\lw}[1]{\smash{\lower1.7ex\hbox{#1}}}
\newcommand{\lww}[1]{\smash{\lower6.7ex\hbox{#1}}}
\newcommand{\bbbar}{\ensuremath{B\overline{B}} }

\newcommand{\btag}{\ensuremath{B_{\rm tag}} }
\newcommand{\bsig}{\ensuremath{B_{\rm sig}} }

\newcommand{\eecl}{\ensuremath{E_{\text{ECL}}} }
\newcommand{\brvalue}{1.54}
\newcommand{\brstaterr}{^{+0.38}_{-0.37}}
\newcommand{\brsysterr}{^{+0.29}_{-0.31}}
\newcommand{\was}{W\c as}
\begin{document}


\preprint{\vbox{ \hbox{   }
    \hbox{Belle Preprint 2010-12}
    \hbox{KEK Preprint 2010-18}
    \hbox{NTLP Preprint 2010-02}
}}

\title{ 
  \quad\\[0.5cm]
  { \bf Evidence for $B^- \to \tau^- \overline{\nu}_{\tau}$ with a Semileptonic Tagging Method\\}
}

\affiliation{Budker Institute of Nuclear Physics, Novosibirsk}
\affiliation{Faculty of Mathematics and Physics, Charles University, Prague}
\affiliation{Chiba University, Chiba}
\affiliation{University of Cincinnati, Cincinnati, Ohio 45221}
\affiliation{The Graduate University for Advanced Studies, Hayama}
\affiliation{Hanyang University, Seoul}
\affiliation{University of Hawaii, Honolulu, Hawaii 96822}
\affiliation{High Energy Accelerator Research Organization (KEK), Tsukuba}
\affiliation{Institute of High Energy Physics, Chinese Academy of Sciences, Beijing}
\affiliation{Institute of High Energy Physics, Vienna}
\affiliation{Institute of High Energy Physics, Protvino}
\affiliation{Institute for Theoretical and Experimental Physics, Moscow}
\affiliation{J. Stefan Institute, Ljubljana}
\affiliation{Kanagawa University, Yokohama}
\affiliation{Institut f\"ur Experimentelle Kernphysik, Karlsruher Institut f\"ur Technologie, Karlsruhe}
\affiliation{Korea Institute of Science and Technology Information, Daejeon}
\affiliation{Korea University, Seoul}
\affiliation{Kyungpook National University, Taegu}
\affiliation{\'Ecole Polytechnique F\'ed\'erale de Lausanne (EPFL), Lausanne}
\affiliation{Faculty of Mathematics and Physics, University of Ljubljana, Ljubljana}
\affiliation{University of Maribor, Maribor}
\affiliation{Max-Planck-Institut f\"ur Physik, M\"unchen}
\affiliation{University of Melbourne, School of Physics, Victoria 3010}
\affiliation{Nagoya University, Nagoya}
\affiliation{Nara Women's University, Nara}
\affiliation{National Central University, Chung-li}
\affiliation{National United University, Miao Li}
\affiliation{Department of Physics, National Taiwan University, Taipei}
\affiliation{H. Niewodniczanski Institute of Nuclear Physics, Krakow}
\affiliation{Nippon Dental University, Niigata}
\affiliation{Niigata University, Niigata}
\affiliation{University of Nova Gorica, Nova Gorica}
\affiliation{Novosibirsk State University, Novosibirsk}
\affiliation{Osaka City University, Osaka}
\affiliation{Panjab University, Chandigarh}
\affiliation{Seoul National University, Seoul}
\affiliation{Shinshu University, Nagano}
\affiliation{Sungkyunkwan University, Suwon}
\affiliation{School of Physics, University of Sydney, NSW 2006}
\affiliation{Tata Institute of Fundamental Research, Mumbai}
\affiliation{Excellence Cluster Universe, Technische Universit\"at M\"unchen, Garching}
\affiliation{Toho University, Funabashi}
\affiliation{Tohoku Gakuin University, Tagajo}
\affiliation{Tohoku University, Sendai}
\affiliation{Department of Physics, University of Tokyo, Tokyo}
\affiliation{Tokyo Metropolitan University, Tokyo}
\affiliation{Tokyo University of Agriculture and Technology, Tokyo}
\affiliation{IPNAS, Virginia Polytechnic Institute and State University, Blacksburg, Virginia 24061}
\affiliation{Yonsei University, Seoul}
 \author{K.~Hara}\affiliation{Nagoya University, Nagoya} 
 \author{T.~Iijima}\affiliation{Nagoya University, Nagoya} 
  \author{H.~Aihara}\affiliation{Department of Physics, University of Tokyo, Tokyo} 
  \author{V.~Aulchenko}\affiliation{Budker Institute of Nuclear Physics, Novosibirsk}\affiliation{Novosibirsk State University, Novosibirsk} 
  \author{T.~Aushev}\affiliation{\'Ecole Polytechnique F\'ed\'erale de Lausanne (EPFL), Lausanne}\affiliation{Institute for Theoretical and Experimental Physics, Moscow} 
  \author{T.~Aziz}\affiliation{Tata Institute of Fundamental Research, Mumbai} 
  \author{A.~M.~Bakich}\affiliation{School of Physics, University of Sydney, NSW 2006} 
  \author{E.~Barberio}\affiliation{University of Melbourne, School of Physics, Victoria 3010} 
  \author{K.~Belous}\affiliation{Institute of High Energy Physics, Protvino} 
  \author{M.~Bischofberger}\affiliation{Nara Women's University, Nara} 
% ...
\collaboration{The Belle Collaboration}

\noaffiliation
  

\begin{abstract}
We present a measurement of the decay $B^{-}\rightarrow\tau^{-}\overline{\nu}_\tau$
using a data sample containing $657\times 10^6$ \bbbar pairs collected at the $\Upsilon(4S)$
resonance with the Belle detector at the KEKB asymmetric-energy $e^{+}e^{-}$ collider.
A sample of $B^+B^-$ pairs are tagged by reconstructing one $B^+$ meson decaying
semileptonically. We detect the $B^-\to \tau^-\overline{\nu}_{\tau}$ candidate in
the recoil.
We obtain a signal with a significance of 3.6 standard deviations including systematic uncertainties,
and measure the branching fraction to be
${\cal B}(B^{-}\rightarrow\tau^{-}\overline{\nu}_{\tau}) =
 [\brvalue\brstaterr(\text{stat})\brsysterr(\text{syst})] \times 10^{-4}$.
This result confirms the evidence for $B^{-}\to\tau^-\overline{\nu}_\tau$ 
obtained in a previous Belle measurement that used a hadronic $B$ tagging method.
\end{abstract}


\maketitle


{\renewcommand{\thefootnote}{\fnsymbol{footnote}}}
\setcounter{footnote}{0}

The purely leptonic decay $B^{-}\rightarrow \tau^{-}\overline{\nu}_\tau$~\cite{conjugate}
is of particular interest since it provides a unique opportunity to test the Standard Model (SM)
and search for new physics beyond the SM.
In the SM, the branching fraction of the decay $B^{-}\rightarrow\tau^{-}\overline{\nu}_\tau$ 
is given by
\begin{equation}
 \label{eq:BR_B_taunu}
{\cal B}(B^{-}\rightarrow\tau^{-}\overline{\nu}_\tau) = \frac{G_{F}^{2}m_{B}m_{\tau}^{2}}{8\pi}\left(1-\frac{m_{\tau}^{2}}{m_{B}^{2}}\right)^{2}f_{B}^{2}|V_{ub}|^{2}\tau_{B},
\end{equation}
where $G_{F}$ is the Fermi coupling constant, $m_{\tau}$ and $m_{B}$ are
the $\tau$ lepton and $B^-$ meson masses,  $\tau_{B}$ is the $B^{-}$ lifetime,
$|V_{ub}|$ is the magnitude of the Cabibbo-Kobayashi-Maskawa (CKM) matrix element~\cite{CKM},
and $f_{B}$ is the $B$ meson decay constant.
Dependence on the lepton mass arises from helicity conservation, which 
suppresses the muon and electron channels.
A recent SM estimation of the branching fraction~\cite{CKMfitter2010} is
$(0.76^{+0.11}_{-0.06})\times 10^{-4}$.
In the absence of new physics, measurement of the $B^{-}\rightarrow\tau^{-}\overline{\nu}_\tau$ decay 
can provide a direct experimental determination of $f_B$, 
which can be compared to lattice QCD calculations~\cite{lattice}.
Physics beyond the SM, however, could significantly suppress
or enhance ${\cal B}(B^{-}\rightarrow\tau^{-}\overline{\nu}_\tau)$ 
via exchange of a new charged particle such as a charged Higgs boson from supersymmetry or 
two-Higgs doublet models~\cite{Hou:1992sy,Baek:1999ch}.

Belle previously reported~\cite{ikado-2006-97} the first evidence of 
 $B^{-}\rightarrow\tau^{-}\overline{\nu}_\tau$ decay
 with a significance of $3.5$ standard deviations ($\sigma$),
 and measured the branching fraction to be
${\cal B}(B^{-}\rightarrow\tau^{-}\overline{\nu}_{\tau}) = 
 (1.79^{+0.56}_{-0.49}(\mbox{stat})^{+0.46}_{-0.51}(\mbox{syst})) \times 10^{-4}$, 
 using a hadronic reconstruction tagging method.
The BaBar Collaboration reported a search for $B^{-}\rightarrow\tau^{-}\overline{\nu}_\tau$
decay with hadronic tagging~\cite{Aubert:2007} using $383 \times 10^6$ \bbbar pairs 
and with semileptonic tagging~\cite{Aubert:2009_semil} using $459 \times 10^6$ \bbbar pairs.
Combining the two measurements, they obtained a 2.8$\sigma$ excess and
a branching fraction
${\cal B}(B^{-}\rightarrow\tau^{-}\overline{\nu}_{\tau})= (1.7\pm0.6)\times 10^{-4}$.
These experimental results are slightly larger than the SM estimation in Ref. \cite{CKMfitter2010},
though the statistical precision is not sufficient to demonstrate 
the existence of physics beyond the SM.
To better establish this decay mode and determine the branching fraction with 
greater precision, we present a measurement of $B^{-}\rightarrow\tau^{-}\overline{\nu}_{\tau}$
from Belle using a semileptonic tagging method.

We use a $605~\textrm{fb}^{-1}$ data sample containing 
$657\times 10^{6}$ \bbbar pairs collected with the Belle detector
at the KEKB asymmetric-energy $e^{+}e^{-}$ ($3.5$ on $8$ GeV) collider~\cite{KEKB}
operating at the $\Upsilon(4S)$ resonance ($\sqrt{s} = 10.58$ GeV).
We also use a data sample of 68~fb$^{-1}$ taken at a center of mass energy
60~MeV below the nominal $\Upsilon(4S)$ mass (off-resonance) for background studies.
The Belle detector~\cite{Belle} is a large-solid-angle magnetic
spectrometer that
consists of a silicon vertex detector (SVD),
a 50-layer central drift chamber (CDC), an array of
aerogel threshold Cherenkov counters (ACC), 
a barrel-like arrangement of time-of-flight
scintillation counters (TOF), and an electromagnetic calorimeter
(ECL) comprised of CsI(Tl) crystals located inside 
a superconducting solenoid coil that provides a 1.5~T
magnetic field.  An iron flux-return located outside of
the coil is instrumented to detect $K_L^0$ mesons and to identify
muons (KLM).  
Two inner detector configurations were used. A 2.0 cm beampipe
and a 3-layer silicon vertex detector were used for the first sample
of 152 $\times 10^6 B\overline{B}$ pairs, while a 1.5 cm beampipe, a 4-layer
silicon detector and a small-cell inner drift chamber were used to record  
the remaining 505 $\times 10^6 B\overline{B}$ pairs \cite{svd2}.  

We use a detailed Monte Carlo (MC) simulation based on GEANT~\cite{GEANT} 
to determine the signal selection efficiency and study the background.
In order to reproduce the effects of beam background, data taken with random
triggers for each run period are overlaid on simulated events. 
The $B^{-}\rightarrow\tau^{-}\overline{\nu}_{\tau}$ signal decay is generated
by the EvtGen package~\cite{EvtGen}.
Radiative effects are modeled using the PHOTOS code~\cite{PHOTOS}.
To model the background from $e^+e^- \to B\overline{B}$ and continuum
$q\overline{q}~(q = u, d, s, c)$ production processes, we use large 
MC samples of $B\overline{B}$ meson pair decays to charm and continuum $q\overline{q}$ processes
corresponding to about ten times and six times the data sample, respectively.
We also use MC samples of rare $B$ decay processes such as charmless 
hadronic, radiative, electroweak decays and $b \to u$ semileptonic decays.
The contamination from other low multiplicity backgrounds
such as $e^+e^-\to \tau^+\tau^-$ and two-photon processes is also 
studied using dedicated MC samples.

The $B^-\to\tau^-\overline{\nu}_\tau$ candidate decays are selected 
using the feature that 
at the $\Upsilon(4S)$ resonance $B$ meson pairs are produced with no additional particles.
We first reconstruct one of the $B$ mesons decaying semileptonically
(referred to hereafter as $\btag$)
and then compare the properties of the remaining particle(s) in the event ($\bsig$) 
to those expected for signal and background.
In order to avoid experimental bias, 
the signal region in data is not examined until the event 
selection criteria are finalized.

Charged particles are selected from well measured tracks 
(reconstructed with the CDC and SVD) originating from the interaction point.
Electron candidates are identified based on a likelihood calculated using
the following information:
$dE/dx$ measured in the CDC, the response of the ACC, the ECL shower shape 
and the ratio of the ECL energy deposited to the track momentum.
Muon candidates are selected using KLM hits associated to a charged track.
Both muons and electrons are selected with efficiency greater than 90\% 
in the momentum region above 1.2 GeV/$c$, and 
misidentification rates of less than 0.2\% (1.5\%) for electrons (muons). 
After selecting leptons, we distinguish charged kaons from pions 
based on a kaon likelihood derived from the TOF, ACC, and $dE/dx$ measurements in the CDC.
The typical kaon identification efficiency is greater than 85\% and 
the probability of misidentifying pions 
as kaons is about 8\%.
Photons are identified as isolated ECL clusters that are not matched to any charged track.
Neutral $\pi^0$ candidates are selected from pairs of photons with invariant mass between 0.118 and
0.150 GeV/$c^2$.
The energy of the photon candidates must exceed: 
50 MeV for the barrel, 100 MeV for the forward endcap and 150 MeV
for the backward endcap, except for low momentum $\pi^0$ candidates from
$\overline{D}^{*0}\to \overline{D}^{0}\pi^0$ decay for which 
we require the photon energy to be greater than 30 MeV.

We reconstruct the $\btag$ in $B^+\to \overline{D}^{*0}\ell^+\nu_\ell$ and
$B^+\to \overline{D}^{0}\ell^+\nu_\ell$ decays, where $\ell$ is electron ($e$) or muon ($\mu$).
$\overline{D}^0$ mesons are reconstructed in the $K^+\pi^-$, $K^+\pi^-\pi^0$ 
and $K^+\pi^-\pi^+\pi^-$ modes.
For $\bsig$, we use $\tau^-$ decays to only one charged particle and neutrinos i.e.
$\tau^- \to \ell^- \overline{\nu}_\ell \nu_\tau$ and
$\tau^- \to \pi^- \nu_\tau$.

We require the invariant mass of $\overline{D}^0$ candidates to be in the range
$[1.851~\text{GeV}/c^2, 1.879~\text{GeV}/c^2]$ for $\overline{D}^0\to K^+\pi^-$ 
and $K^+\pi^-\pi^+\pi^-$ decays,
and $[1.829~\text{GeV}/c^2, 1.901$~GeV/$c^2]$ for $\overline{D}^0\to K^+\pi^-\pi^0$ decay.
$\overline{D}^{*0}$ candidates are selected by combining the $\overline{D}^0$ candidates 
with low momentum $\pi^0$ candidates or photons.
For $\overline{D}^{*0}$ candidates, we require the mass difference
$\Delta M \equiv M_{D^{*0}} - M_{D^0}$ to be in the range 
$[0.1389~\text{GeV}/c^2, 0.1455~\text{GeV}/c^2]$
and $[0.123~\text{GeV}/c^2, 0.165~\text{GeV}/c^2]$ for
$\overline{D}^{*0}\to \overline{D}^0\pi^0$ and $\overline{D}^{*0}\to \overline{D}^0\gamma$ decays, 
respectively.
These regions correspond to three standard deviations in the corresponding resolutions.
To suppress $\overline{D}^{*0}$'s from continuum background processes, the momentum of $\overline{D}^{*0}$ 
candidates calculated in the $\Upsilon(4S)$ center-of-mass system (cms) is required 
to be less than 2.5 GeV/$c$.

\begin{figure}
\begin{center}
        \includegraphics[width=0.45\textwidth]{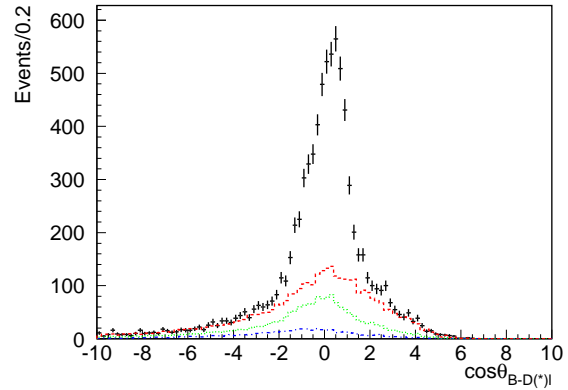}
    \caption{
      $\cos\theta_{B,D^{(*)}\ell}$ distribution for candidate events with $\eecl < 1.2$ GeV
      selected with all $\btag$ and $\bsig$ requirements except for 
      those on $\cos\theta_{B,D^{(*)}\ell}$.
      Leptonic and hadronic $\tau$ decay modes are combined.
      The points with error bars are data. The dot-dashed, dotted and dashed histograms are
      the MC expectation for events without $B^+ \to \overline{D}^{(*)0} \ell^+ \nu_\ell$
      decays for $B^+B^-$, sum of $B^+B^-$ and $B^0\bar{B}^0$,
      and sum of \bbbar and non-\bbbar events, respectively.
      }
    \label{fig:cosbdl}
\end{center}
\end{figure}

We select signal candidates from events with one $\overline{D}^0$ or $\overline{D}^{*0}$
and one $\ell^+$ to form $\btag$, and one $\ell^-$ or $\pi^-$ candidate for $\bsig$.
We require that no other charged particle or $\pi^0$ remain in the event after removing
the particles from the $\btag$ and $\bsig$ candidates.
The $\btag$ candidates are selected using the lepton momentum in the cms frame,
$P^{\text{cms}}_\ell$, and the cosine of the angle between
the direction of the $\btag$ momentum and the direction of the momentum sum of the 
$\overline{D}^{(*)0}$ and the lepton, $\cos\theta_{B,D^{(*)}\ell}$, measured in the cms frame.
This angle is calculated using
\begin{equation}
    \cos\theta_{B,D^{(*)}\ell} = \frac{2E^{\text{cms}}_{\text{beam}}E^{\text{cms}}_{D^{(*)}\ell}-m^2_B-M^2_{D^{(*)}\ell}}{2 P^{\text{cms}}_B\cdot P^{\text{cms}}_{D^{(*)}\ell}},
\end{equation}
where $E^{\text{cms}}_{\text{beam}}$ is the beam energy,
$P^{\text{cms}}_B$ is the momentum of $B$ meson calculated with $P^{\text{cms}}_B = \sqrt{(E^{\text{cms}}_{\text{beam}})^2 - m_B^2}$,
$E^{\text{cms}}_{D^{(*)}\ell}$, $P^{\text{cms}}_{D^{(*)}\ell}$ and $M_{D^{(*)}\ell}$ are the energy sum, momentum sum 
and invariant mass, respectively, of the $\overline{D}^{(*)0}$ and lepton system.
All parameters are calculated in the cms. Properly reconstructed $\btag$ candidates 
are populated within the physical range $[-1,1]$, 
while combinatorial backgrounds can take unphysical values.
For the signal side, the $\ell^-$ or $\pi^-$ candidate from the $\tau$ decay
is selected using the momentum in the cms, denoted $P^{\text{cms}}_{\rm sig}$.
The signal yield is obtained by fitting the distribution of the 
remaining energy in the ECL, denoted $\eecl$, which is the sum of
the energies of ECL clusters that are not associated with particles 
from the $\btag$ and $\bsig$ candidates;
here the $\eecl$ clusters satisfy the same minimum energy requirements
as photon candidates.
For signal events, $\eecl$ must be either zero or a small value
arising from splitoff showers created by $\btag$ and $\bsig$ particles
and residual beam background hits. Therefore, signal events peak at low $\eecl$.
On the other hand, background events are distributed toward higher 
$\eecl$ due to the contribution from additional particles.
The selection criteria for $P^{\text{cms}}_\ell$, $\cos\theta_{B,D^{(*)}\ell}$ and $P^{\text{cms}}_{\rm sig}$ 
are optimized for each of the $\tau$ decay modes,
because the background levels and the background components are mode-dependent.
The optimization is done so that the figure of merit $s/\sqrt{s+n}$ is maximized,
where $s$ and $n$ are the number of signal and background events 
expected in the signal-enhanced region $\eecl < 0.2$~GeV,
calculated assuming a signal branching fraction of $1.79 \times 10^{-4}$.
For leptonic $\tau$ decays, the dominant background is from \bbbar events
tagged by a semileptonic decay with a correctly reconstructed combination 
of a $\overline{D}^{*0}$ and a $\ell^+$.
For these decays loose selection criteria are chosen to maintain high signal efficiency:
$0.5~\text{GeV}/c < P^{\text{cms}}_{\ell} < 2.5$ GeV/$c$, $-2.1 < \cos\theta_{B,D^{*}\ell}<1.3$ 
for the $\overline{D}^{*0}$ mode or $-2.6 < \cos\theta_{B,D\ell}<1.2$ for the 
$\overline{D}^0$ mode, and $0.3 \text{ GeV}/c < P^{\text{cms}}_{\rm sig}$.
For the hadronic $\tau$ decay mode, there is more background 
from $e^+e^- \to q\overline{q}$ continuum and combinatorial $D^{(*)0}\ell$ background.
Tighter criteria are used to reduce such backgrounds:
$1.0 \text{ GeV}/c < P^{\text{cms}}_{\ell} < 2.2$ GeV/$c$, $-1.1 < \cos\theta_{B,D^{(*)0}\ell}<1.1$,
and $1.0 \text{ GeV}/c < P^{\text{cms}}_{\rm sig} < 2.4 $ GeV/$c$.
The upper bound on $P^{\text{cms}}_{\rm sig}$ is introduced to reject two-body $B$ decays.
In addition, we suppress continuum background by requiring the cosine of the angle 
between the signal side pion track and the thrust axis of the $\btag$, 
$\cos\theta_{ \text{thr} }$, to be less than 0.9.
We select candidate events in the range $\eecl < 1.2$ GeV for further analysis.
The number of candidate events are 2481 for $\tau^- \to e^- \overline{\nu}_e \nu_\tau$,
2011 for $\tau^- \to \mu^- \overline{\nu}_\mu \nu_\tau$ and
1018 for $\tau^- \to \pi^- \nu_\tau$ decays.
Figure~\ref{fig:cosbdl} shows the $\cos\theta_{B,D^{(*)}\ell}$ distribution for the 
signal candidate events including both leptonic and hadronic $\tau$ decay modes
with all selection criteria other than $\cos\theta_{B,D^{(*)}\ell}$ applied.
The excess over the MC expectation for events without 
$B \to D^{(*)} \ell \nu$ decays indicates that the final sample contains candidate events 
with the correct combination of a $\overline{D}^{(*)0}$ and a $\ell^+$ forming a $\btag$.
In the remaining candidates, according to a MC study,
4.6\%, 13.4\% and 12.0\% are events without a $\btag$ from $B^+B^-$, $B^0\bar{B}^0$ and
non-\bbbar processes, respectively.


The number of signal events is extracted from an extended maximum likelihood fit
to the \eecl distribution of the candidate events.
Probability density functions (PDFs) for each $\tau$ decay mode
are constructed from the MC simulation.
We use \eecl histograms obtained from MC samples for each of the signal and
the background components. 
The PDFs are combined into a likelihood function,
\begin{equation}
{\cal L} = \frac{e^{-\sum_j n_j}}{N!}
\prod_{i=1}^{N}\sum_j n_j f_j(E_{i})
\end{equation}
where $j$ is an index for the signal and background contributions, 
$n_{j}$ and $f_{j}$ are the yield and the PDF, respectively, of the $j$th component,
$E_{i}$ is the $\eecl$ value in the $i$th event, and 
$N$ is the total number of events in the data.
The dominant background components are from \bbbar decays to a final state 
with charm and continuum processes.
The small background from rare charmless $B$ decays
and other low multiplicity processes such as $\tau$ pair and two-photon processes
is also included in the fit.
In the final sample with $\eecl < 1.2$ GeV, the fractions of the background from rare 
charmless $B$ decays and low multiplicity non-$B$ processes
are estimated from MC to be 8\% and 3\% for leptonic $\tau$ decays and 11\% and 8\% for 
hadronic $\tau$ decay, respectively.

\begin{figure}
\begin{center}
        \includegraphics[width=0.4\textwidth]{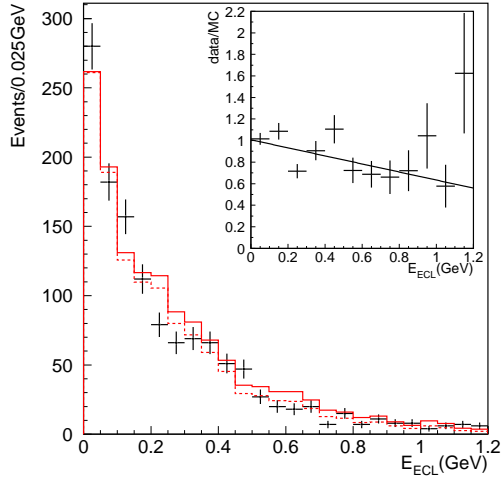}
    \caption{
      \eecl distribution for double semileptonic tagged events.
      The points with error bars are data and the solid histogram is the MC expectation scaled 
      to the luminosity of the data.
      The dashed histogram is the MC expectation multiplied by the correction function
      described in the text.
      The inset shows the ratio of data to the MC expectation and the correction function. 
    }
    \label{fig:doubletag}
\end{center}
\end{figure}

The $\eecl$ estimation in MC is validated using various control samples.
The MC distributions of not only $\eecl$ but also $P^{\text{cms}}_{\ell}$, 
$\cos\theta_{B,D^{(*)}\ell}$, $P^{\text{cms}}_{\rm sig}$ and 
$\cos\theta_{ \text{thr} }$ are compared to those of the control samples to confirm
that MC describes the background composition properly.
The off-resonance data is used to calibrate the MC simulation of the continuum
background.
We find that our MC underestimates the overall normalization of the continuum background 
though the predicted shapes are consistent with data within statistical errors.
We obtain the correction factor for the overall normalization of the continuum MC 
to be $1.43 \pm 0.11$ by comparing the number of remaining events in off-resonance 
data with the MC expectation.
The sidebands in $\cos\theta_{B,D^{(*)}\ell}$, $\overline{D}^0$ mass,
the mass difference between $\overline{D}^{*0}$ and $\overline{D}^0$, and $\eecl$ 
are used as control samples to check the overall background description including 
the $B\overline{B}$ contribution.
The distributions in these variables obtained from MC
with the continuum normalization correction applied
are found to be consistent with the corresponding distributions in data.
The agreement between MC and data is also confirmed in $B^0$ tagged
events where the $\btag$ is reconstructed in $B^0 \to D^{*-}\ell^+\nu$ decays.
The contributions to the $\eecl$ distribution are not only from beam background 
but also include splitoff showers originating from $\btag$ and $\bsig$ decay products.
The relative fractions of these sources are 21\%, 53\% and 26\%, respectively,
in the signal MC sample.
To take into account the possible difference between MC and data descriptions of 
splitoff showers, the signal $\eecl$ shape is calibrated using double tagged events,
in which the $B_{\rm tag}$ is reconstructed
in a semileptonic decay as described above and $B_{\rm sig}$ is reconstructed in the decay chain,
$B^{-} \rightarrow D^{*0}\ell^{-}\overline{\nu}$ ($D^{*0}\rightarrow D^{0}\pi^{0}$),
followed by $D^0 \to K^- \pi^+$.
Figure~\ref{fig:doubletag} shows the $\eecl$ distribution in the
double tagged sample for data and for the MC simulation scaled to the same luminosity.
The background in this control sample is negligibly small.
We find the \eecl distribution of data tends to have a slightly smaller width than MC.
The difference between the data and MC is parameterized as a first-order polynomial
function of \eecl obtained by fitting the ratio of data to MC for the \eecl histograms
of the double tagged sample.
The ratio and the fit result are also shown in Fig.~\ref{fig:doubletag}.
The \eecl histogram obtained from the signal MC sample is multiplied by 
this correction function.

\begin{figure}[bh]
    \begin{center}
	\includegraphics[width=0.5\textwidth]{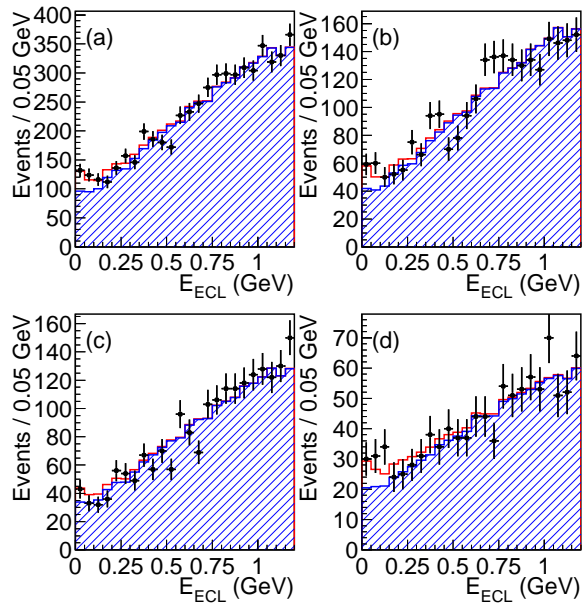}	
	\caption{\eecl distribution of semileptonic tagged events
	  with the fit result for 
	  (a) all $\tau$ decay modes combined,
	  (b) $\tau^- \to e^-\overline{\nu}_e\nu_\tau$,
	  (c) $\tau^- \to \mu^-\overline{\nu}_\mu\nu_\tau$ and
	  (d) $\tau^- \to \pi^-\nu_\tau$.
	  The points with error bars are data. The hatched histogram
	  and solid open histogram are the background and the signal contributions,
          respectively.
	  }
	\label{fig:eecl_fit}
    \end{center}
\end{figure}

In the final fit, four parameters are allowed to vary: the total signal yield
and the sum of \bbbar and continuum backgrounds for each $\tau$ decay mode.
The ratio of the \bbbar to the continuum background is fixed to the value obtained from MC
with the normalization correction applied.
Other background contributions are fixed to the MC expectation.
We combine $\tau$ decay modes by constraining the ratios of the signal 
yields to the ratio of the reconstruction efficiencies obtained from MC
including the branching fractions of $\tau$ decays~\cite{pdg2010}.
Figure~\ref{fig:eecl_fit} shows the $\eecl$ distribution overlaid with the fit results.
The \eecl distribution for each $\tau$ decay mode is also shown.
We see a clear excess of signal events in the region near zero
and obtain a signal yield of $n_{\rm s} = 143^{+36}_{-35}$.
The branching fraction is calculated as
${\cal B} = n_{\rm s}/(2\varepsilon N_{B^{+}B^{-}})$,
where $\varepsilon$ is the reconstruction efficiency including 
the branching fraction of the $\tau$ decay mode and 
$N_{B^{+}B^{-}}$ is the number of $\Upsilon(4S)\rightarrow B^{+}B^{-}$ 
events, assuming $N_{B^{+}B^{-}} = N_{B^{0}\overline{B}^{0}}$.
Table~\ref{tab:fit_result} lists the signal yields and the branching fractions
obtained from separate fits to each $\tau$ decay mode and the fit with all 
three modes combined. 
The results of the individual fits are consistent within statistics.
The $\chi^2$ of the three results is 2.43
for two degrees of freedom, corresponding to a $30$\% confidence level.

\begin{table}
 \begin{center}
    \caption{
      Results of the fit for signal yields and branching fractions.
      $\varepsilon$ is the reconstruction efficiency including 
      the branching fraction of the $\tau$ decay mode.
      The first error in the branching fraction is statistical and the second is 
      systematic.
      }
   \label{tab:fit_result}
    \begin{tabular}{lcccc} \hline \hline
Decay Mode & Signal Yield~ &  $\varepsilon$, $10^{-4}$ & ${\cal B}$, $10^{-4}$ \\ \hline
$\tau^-\to e^{-}\overline{\nu}_e \nu_{\tau}$     & $73^{+23}_{-22}$  & $5.9$  & $1.90^{+0.59}_{-0.57}{}^{+0.33}_{-0.35}$ \\
$\tau^-\to \mu^{-}\overline{\nu}_\mu \nu_{\tau}$ & $12^{+18}_{-17}$  & $3.7$  & $0.50^{+0.76}_{-0.72}{}^{+0.18}_{-0.21}$ \\
$\tau^-\to\pi^{-}\nu_{\tau}$                     & $55^{+21}_{-20}$  & $4.7$  & $1.80^{+0.69}_{-0.66}{}^{+0.36}_{-0.37}$ \\
\hline
Combined                                         & $143^{+36}_{-35}$ & $14.3$ & $1.54^{+0.38}_{-0.37}{}^{+0.29}_{-0.31}$ \\
\hline\hline
    \end{tabular}
 \end{center}
\end{table}

Systematic errors for the measured branching fraction are associated with 
the uncertainties in the signal yield, efficiencies and the number of $B^{+}B^{-}$ pairs.
Unless explicitly stated otherwise, the systematic errors for each source are
obtained by varying the corresponding parameters individually by their
uncertainties, repeating the fit procedure and adding differences from the nominal
result in quadrature.
The systematic errors for the signal yield arise from the uncertainties in the PDF shapes
for the signal and for the background.
The uncertainty in the signal shape correction function is estimated by changing the 
parameters of the correction function by their errors and replacing the function with 
a second-order polynomial ($^{+1.9}_{-2.4}$\%). 
The systematic error from MC statistics is evaluated by 
varying the content of each bin in the signal $\eecl$ PDF histograms by its statistical 
uncertainty ($\pm0.9$\%).
The main contributions to the systematic errors for the background PDF shapes are
statistical errors in the MC histograms ($^{+8.6}_{-8.3}$\%), which is estimated in the same 
way as the signal PDF MC statistical uncertainty.
Other large sources are the uncertainties in the background composition.
The errors due to the uncertainties in the branching fractions of $B$ decay modes
that peak near zero $\eecl$ such as $B^-\to D^0\ell^-\overline{\nu}_\ell$ 
with $D^0\to K_L^0 K_L^0$, $K_L^0 \pi^0$ and $K^- \ell^+\nu_\ell$,
and $\overline{B}^0 \to D^+ \ell^- \overline{\nu}_\ell$ with $D^+ \to 
K_L^0 \ell^+ \nu_\ell$ are estimated by changing the branching fractions in 
MC by their errors~\cite{pdg2010} ($^{+4.5}_{-8.8}$\%).
For branching fractions of $D$ decays with a $K^0_L$,
we use the values for the corresponding $D$ decays with $K^0_S$'s.
Uncertainties in the background from the possible contribution of rare charmless
$B$ decays such as $B^- \to \pi^0 \ell^- \overline{\nu}_\ell$, $B^- \to K^-\nu\overline{\nu}$ and
$\ell^- \overline{\nu}_\ell \gamma$, and from $\tau^+\tau^-$ pair and two photon processes
are evaluated by changing the fractions obtained from MC by their experimental 
errors~\cite{pdg2010} if available, or by $\pm50$\% otherwise ($^{+7.6}_{-7.7}$\%).
The systematic error due to the uncertainty in the normalization correction factor for the 
continuum MC is $^{+2.6}_{-2.5}$\%.
The systematic error associated with the reconstruction 
efficiency of the tag-side $B$ is evaluated by comparing
of the ${\cal B}(B^{-}\rightarrow D^{*0}\ell^{-}\overline{\nu}_\ell)$ branching fraction
measured with the double tagged sample in data to the world average value~\cite{pdg2010}.
We obtain the ratio to be $0.907\pm0.044$ and take the difference from unity plus one $\sigma$ 
as the systematic error (13.7\%).
The systematic errors in the signal-side efficiencies arise from the uncertainty 
in tracking efficiency (1.0\%), particle identification efficiency (1.3\%),
branching fractions of $\tau$ decays (0.4\%), and MC statistics (0.8\%).
The systematic error due to the uncertainty in $N_{B^{+}B^{-}}$ is 1.4\%.
The total fractional systematic uncertainty is $^{+19}_{-20}\%$,
and the branching fraction is
\begin{equation}
{\cal B}(B^{-}\rightarrow\tau^{-}\overline{\nu}_{\tau}) = (\brvalue\brstaterr(\text{stat})\brsysterr(\text{syst}))\times 10^{-4}.
\end{equation}
The significance of the observed signal is evaluated by 
$\Sigma = \sqrt{-2\ln({\cal L}_{0}/{\cal L}_{\rm max})}$ 
where ${\cal L}_{\rm max}$ and ${\cal L}_{0}$ denote the maximum likelihood 
value and likelihood value obtained assuming zero signal events, respectively.
The systematic uncertainty is convolved in the likelihood with a Gaussian distribution 
having a width corresponding to the systematic error of the signal yield.
We find the significance of the signal yield to be 3.6$\sigma$.

In summary, we have measured
the decay $B^{-}\rightarrow\tau^{-}\overline{\nu}_\tau$ with \bbbar events tagged by 
semileptonic $B$ decays
using a data sample containing $657\times 10^6$ \bbbar pairs collected at the $\Upsilon(4S)$
resonance with the Belle detector at the KEKB asymmetric-energy $e^{+}e^{-}$ collider. 
We measure the branching fraction to be 
$(\brvalue\brstaterr(\text{stat})\brsysterr(\text{syst}))\times 10^{-4}$,
with a significance of 3.6 standard deviations including systematics.
This result is consistent with the previous Belle measurement
using \bbbar events tagged by hadronic $B$ decays and is consistent with the results 
reported by the BaBar collaboration.
Using the measured branching fraction and known values of $G_F$, $m_B$, 
$m_{\tau}$ and $\tau_B$~\cite{pdg2010}, the product
of the $B$ meson decay constant $f_B$ and the magnitude of the 
Cabibbo-Kobayashi-Maskawa matrix element $|V_{ub}|$ 
is determined to be
\begin{equation}
f_B |V_{ub}|= (9.3^{+1.2}_{-1.1}\pm0.9) \times 10^{-4}~\text{GeV}.
\end{equation}
Using $|V_{ub}|=(3.89\pm0.44)\times 10^{-3}$ in Ref.~\cite{pdg2010}, $f_B$ is 
calculated to be $0.24\pm0.05~\text{GeV}$.
The measured branching fraction is consistent within errors with the SM expectation 
from other experimental constraints~\cite{CKMfitter2010}.
The result can be used to extract constraints on new physics models.

\begin{acknowledgements}
We thank the KEKB group for excellent operation of the
accelerator, the KEK cryogenics group for efficient solenoid
operations, and the KEK computer group and
the NII for valuable computing and SINET3 network support.  
We acknowledge support from MEXT, JSPS and Nagoya's TLPRC (Japan);
ARC and DIISR (Australia); NSFC (China); MSMT (Czechia);
DST (India); MEST, NRF, NSDC of KISTI, and WCU (Korea); MNiSW (Poland); 
MES and RFAAE (Russia); ARRS (Slovenia); SNSF (Switzerland); 
NSC and MOE (Taiwan); and DOE (USA).
\end{acknowledgements}


\begin{thebibliography}{99}
\bibitem{conjugate}
The charge-conjugate decays are implied throughout this paper unless otherwise stated.
\end{thebibliography}
\end{document}